\documentclass[iop]{emulateapj}

\usepackage{url}
\usepackage[utf8]{inputenc}
\usepackage{fancyref}
\usepackage{hyperref}
\hypersetup{colorlinks,breaklinks,
            urlcolor=[rgb]{0.2,0.2,0.2},  
            linkcolor=[rgb]{0.843,0.098,0.110},  
            citecolor=[rgb]{0,0.4,1.0}}  
\usepackage{amssymb,amsmath}
\usepackage{enumitem}
\usepackage{natbib}
\usepackage{epstopdf}
\usepackage{rotating}
\usepackage{color}
\usepackage{float}
\restylefloat{table}
\usepackage{graphicx}

\newcommand{\teff}{${T}_{\mathrm{eff}}$}
\newcommand{\logg}{$\log{g}$}
\newcommand{\msun}{$M_{\odot}$}
\newcommand{\rsun}{$R_{\odot}$}

\slugcomment{Accepted to ApJ}

\shorttitle{Pruning The ELM Survey}
\shortauthors{Bell et al.}

\begin{document}

\title{Pruning The ELM Survey: Characterizing Candidate Low-Mass White Dwarfs\\ Through Photometric Variability}
\author{Keaton~J.~Bell\altaffilmark{1,2}, A.~Gianninas\altaffilmark{3}, J.~J.~Hermes\altaffilmark{4,5}, D.~E.~Winget\altaffilmark{1,2}, Mukremin Kilic\altaffilmark{3},  M.~H.~Montgomery\altaffilmark{1,2}, B.~G.~Castanheira\altaffilmark{1,2,6},  Z.~Vanderbosch\altaffilmark{1,2}, K.~I.~Winget\altaffilmark{1,2}, and Warren~R.~Brown\altaffilmark{7}}
\altaffiltext{1}{Department of Astronomy, University of Texas at Austin, Austin, TX\,-\,78712, USA}
\altaffiltext{2}{McDonald Observatory, Fort Davis, TX\,-\,79734, USA}
\altaffiltext{3}{Homer L. Dodge Department of Physics and Astronomy, University of Oklahoma, Norman, OK\,-\,73019, USA}
\altaffiltext{4}{Department of Physics and Astronomy, University of North Carolina, Chapel Hill, NC\,-\,27599, USA}
\altaffiltext{5}{Hubble Fellow}
\altaffiltext{6}{Baylor University, Waco, TX\,-\,76798, USA}
\altaffiltext{7}{Smithsonian Astrophysical Observatory, Cambridge, MA\,-\,02138, USA}

\email{keatonb@astro.as.utexas.edu}

\begin{abstract}
We assess the photometric variability of nine stars with spectroscopic \teff\ and \logg\ values from the ELM Survey that locate them near the empirical extremely low-mass (ELM) white dwarf instability strip. We discover three new pulsating stars: SDSS\,J135512.34+195645.4, SDSS\,J173521.69+213440.6 and SDSS\,J213907.42+222708.9.  However, these are among the few ELM Survey objects that do not show radial velocity variations to confirm the binary nature expected of helium-core white dwarfs. The dominant 4.31-hr pulsation in SDSS\,J135512.34+195645.4 far exceeds the theoretical cutoff for surface reflection in a white dwarf, and this target is likely a high-amplitude $\delta$ Scuti pulsator with an overestimated surface gravity.  We estimate the probability to be less than 0.0008 that the lack of measured radial velocity variations in four of eight other pulsating candidate ELM white dwarfs could be due to low orbital inclination.  Two other targets exhibit variability as photometric binaries.  Partial coverage of the 19.342-hr orbit of WD\,J030818.19+514011.5 reveals deep eclipses that imply a primary radius $>0.4$~\rsun ---too large to be consistent with an ELM white dwarf. The only object for which our time series photometry adds support to the ELM white dwarf classification is SDSS\,J105435.78$-$212155.9, with consistent signatures of Doppler beaming and ellipsoidal variations.  We interpret that the ELM Survey contains multiple false positives from another stellar population at ${T}_{\mathrm{eff}} \lesssim 9{,}000$ K, possibly related to the sdA stars recently reported from SDSS spectra.

\end{abstract}


\section{Introduction}

The Galaxy is not old enough for $\lesssim$\,$0.3$ \msun\ white dwarfs (WDs) to have formed in isolation, even from high-metallicity systems \citep{Kilic2007}. These objects are instead formed as the remnants of mass transfer in post-main-sequence common-envelope binaries.  A close companion can strip away material if a star overflows its Roche lobe while ascending the red giant branch, leaving behind an extremely low-mass (ELM) WD with a degenerate helium core and hydrogen-dominated atmosphere \citep[e.g.,][]{Nelemans2001}. \citet{Althaus2013} and \citet{Istrate2016} have calculated the most recent evolutionary ELM WD models, and \citet[][Section 8]{Heber2016} provides a nice overview of these objects.

The ELM Survey \citep{ELM1,ELM3,ELM5,ELM7,ELM2,ELM4,ELM6} is a spectroscopic effort to discover and characterize ELM WDs.  So far, this survey has measured spectra of 88 objects with parameters from line profiles consistent with He-core WDs ($5.0 \lesssim$ \logg\ $\lesssim 7.0$ and 8000 K $\lesssim$ \teff\ $\lesssim$ 22{,}000 K). Membership to close ($P_{\rm orb} < 25$\,hr) binary systems through measured radial velocity (RV) variations supports the mass-transfer formation scenario for 76 targets.

\tabcolsep=0.15cm
\begin{deluxetable*}{l c c c c c c c c c}[t]
\tablecolumns{9}
\tablecaption{Target Physical Parameters from The ELM Survey\label{tab:objs}}
\tablehead{
\colhead{SDSS} & \colhead{R.A.} & \colhead{Dec.}	& \colhead{\teff} &  \colhead{\logg}  &  \colhead{$M_1$}  &  \colhead{$g_0$} & \colhead{$P_{\rm orb}$} &  \colhead{$K_1$} &  \colhead{Ref.} \\  \colhead{} & \colhead{(h:m:s)} & \colhead{(d:m:s)}	& \colhead{(K)} &  \colhead{(cm s$^{-1}$)}  &  \colhead{(\msun)}  &  \colhead{(mag)} &  \colhead{(days)} &  \colhead{(km s$^{-1}$)} & \colhead{(ELM)}  }
\startdata
J0308+5140  & 03:08:18.19 & +51:40:11.5 & $8380(140) $ & $5.51(0.10)$ & $0.151(0.024)$ & $13.05(0.01)$  & $ 0.8059(0.0004)$ & 78.9(2.7) &6 \\
J1054$-$2121  & 10:54:35.78 & $-$21:21:55.9 & $9210(140) $ & $6.14(0.13)$ & $0.168(0.011)$ & $18.49(0.01)$  & $ 0.104(0.007)$ & 261.1(7.1) &6\\
J1108+1512  & 11:08:15.51 & +15:12:46.7 & $8700(130) $ & $6.23(0.06)$ & $0.167(0.010)$ & $18.83(0.02)$  & $  0.123(0.009)$ & 256.2(3.7) &6 \\
J1449+1717 & 14:49:57.15 & +17:17:29.3 & $ 9700(150)$ & $6.08(0.05)$ & $ 0.168(0.010)$ & $17.62(0.02)$  & $ 0.29075(0.00001)$ &  228.5(3.2) & 6 \\
J1017+1217 & 10:17:07.11 & +12:17:57.4 & $8330(130) $ & $5.53(0.06)$ & $0.142(0.012)$ & $17.48(0.02)$  & $\ldots$ & $<30.2$ & 7 \\
J1355+1956  &  13:55:12.34 & +19:56:45.4 & $8050(120)$ & $ 6.10(0.06)$ & $0.156(0.010)$ & $16.10(0.02)$  & $\ldots$ & $<40.9$ &7 \\
J1518+1354  & 15:18:02.57 & +13:54:32.0 & $8080(120) $ & $ 5.44(0.07)$ & $0.147(0.018)$ & $18.99(0.02)$  & $ 0.577(0.007)$ &  112.7(4.6) &7 \\
J1735+2134  & 17:35:21.69 & +21:34:40.6 & $7940(130) $ & $5.76(0.08)$ & $0.142(0.010)$ & $15.90(0.01)$  & $\ldots$ & $<31.6 $&7 \\
J2139+2227  &  21:39:07.42 & +22:27:08.9 & $7990(130)$ & $5.93(0.12)$ & $0.149(0.011)$ & $15.60(0.01)$  & $\ldots$ & $<22.0$& 7 
\enddata
\end{deluxetable*}

The reliability of spectral line profiles as an ELM diagnostic is challenged by the discovery of thousands of objects that exhibit spectra consistent with low-\logg\ WD models with ${T}_{\mathrm{eff}} \lesssim 9{,}000$ K in recent Sloan Digital Sky Survey (SDSS) data releases \citep{Kepler2016}.  The nature of this large population is under debate, as different observational aspects weigh for-or-against different physical interpretations, including ELM WDs, main sequence A stars in the Galactic halo, or binaries comprised of a subdwarf and a main sequence F, G, or K dwarf \citep{Pelisoli2016}.  These objects are labeled as ``sdA'' stars, with the ELM classification reserved only for those with supporting orbital parameters from RV variations. Only $\approx$15\% of ELM Survey objects are found to have ${T}_{\mathrm{eff}} < 9{,}000$ K.

Six pulsating stars have been published as ELM WDs in a low-mass extension of the hydrogen-atmosphere (DA) WD instability strip from time series photometry obtained at McDonald Observatory \citep{ELMV1, ELMV23, ELMV45, ELMV6}.  However, only the first three discovered show RV variations in available time series spectroscopy. Another pulsating ELM variable (ELMV) in a binary system with a millisecond pulsar was reported by \citet{ELMV7}. Stellar pulsations in these objects provide the potential to constrain the details of their interior structures and to better understand their formation histories through asteroseismology.  The pulsational properties of ELM WDs have been explored theoretically by \citet{VanGrootel2013} and \citet{Corsico2014,Corsico2016a}. The DA WD instability strip is both empirically and theoretically found to shift to lower \teff\ with lower \logg , intersecting the population of sdAs in the ELM regime.

ELM WDs can also exhibit photometric variability that results from their binary nature, including signatures of eclipses, ellipsoidal variations (tidal distortions), and relativistic Doppler beaming \citep[also called Doppler boosting;][]{Shporer2010,Kilic2011,Hermes2014}. In the case of the 12.75-minute binary SDSS J0651+2844, these have enabled the measurement of orbital decay from gravitational radiation \citep{J0651}.

In addition to these variables, numerous stars have been published as pulsating precursors to ELM WDs (pre-ELMs). \citet{Maxted2013,Maxted2014} discovered two recently stripped cores of red giants that pulsate in binary systems with main sequence A stars.  \citet{Corti2016} reported on two variable stars that occupy a region of parameter space where they could plausibly be either pre-ELM WD or SX Phoenicis pulsators.
Finally, \citet{Gianninas2016} discovered three pre-ELMs with mixed H/He atmospheres that pulsate at higher temperatures than an extrapolation of the empirical DA WD instability strip due to the presence of He in their atmospheres.  \citet{Corsico2016b} have explored the properties of pre-ELM WD pulsations in the evolutionary models of \citet{Althaus2013}.

In this work, we assess the photometric variability of nine candidate ELM WD pulsators from The ELM Survey papers VI \citep[ELM6;][]{ELM6} and VII \citep[ELM7;][]{ELM7}.  We describe our candidate selection and observations in Section~\ref{sec:obs}.  We present an object-by-object analysis in Section~\ref{sec:anal}.  We discuss our new variable and non-variable objects in the context of the rapidly developing picture of ELM WD parameter space in Section~\ref{sec:disc} and conclude with a summary in Section~\ref{sec:conc}.

\section{Observations}
\label{sec:obs}

Our observing campaign targeted all nine stars published in the ELM6 and ELM7 samples with \logg\ $< 7.0$ and \teff\ within 500 K of the current empirical ELMV instability strip, which has been updated to reflect the spectroscopic corrections derived from 3D convection models in the ELM regime \citep{Tremblay2015, ELM6}.  Select physical parameters published for these stars by the ELM Survey are listed in Table~\ref{tab:objs}.

\begin{deluxetable}{l r c c c r}[b]
\tablecolumns{9}
\tablecaption{Journal of Observations \label{tab:obs}}
\tablehead{
\colhead{SDSS} & \colhead{Date} & \colhead{Exposure} & \colhead{Run Duration} \\
\colhead{} & \colhead{(UTC)} & \colhead{Time (s)}	& \colhead{(h)} }
\startdata
J0308+5140    & 11 Oct 2015 & 3 & 4.9 \\
                        & 12 Oct 2015 & 3 & 1.7  \\
                        & 13 Oct 2015 & 3 & 2.7  \\
                        & 06 Feb 2016 & 3 & 4.1  \\
J1054$-$2121 & 15 Mar 2015 & 20 & 1.7  \\
                        & 20 Apr 2015 & 30 & 3.0  \\
                        & 21 Apr 2015 & 20 & 4.2  \\
J1108+1512    & 19 Mar 2015 & 30 & 0.9  \\
                        & 12 Mar 2016 & 30 & 1.6  \\
                        & 12 Mar 2016 & 60 & 4.0  \\
                        & 16 Mar 2016 & 30 & 4.3  \\
                        & 01 May 2016 & 15 & 2.5  \\
J1449+1717    & 23 Jul 2014 & 15 & 2.3  \\
                        & 24 Jul 2014 & 25 & 2.6  \\
                        & 14 Apr 2016 & 5 & 0.6  \\
                        & 14 Apr 2016 & 15 & 2.9  \\
J1017+1217    & 08 Jan 2016 & 5 & 2.2  \\
                        & 09 Jan 2016 & 30 & 3.5  \\
                        & 11 Mar 2016 & 5 & 3.9  \\
                        & 17 Mar 2016 & 10 & 3.4  \\
                        & 30 Apr 2016 & 5 & 2.1  \\
                        & 03 May 2016 & 10 & 3.9  \\
J1355+1956    & 14 Apr 2016 & 3 & 2.6  \\
                        & 04 May 2016 & 3 & 1.5  \\
                        & 05 May 2016 & 3 & 2.0  \\
                        & 06 May 2016 & 5 & 6.4  \\
J1518+1354    & 15 Apr 2016 & 30 & 4.3  \\
J1735+2134    & 30 Apr 2016 & 3 & 4.5  \\
                        & 01 May 2016 & 3 & 0.9  \\
                        & 01 May 2016 & 3 & 3.0  \\
                        & 03 May 2016 & 3 & 4.1  \\
                        & 07 May 2016 & 5 & 2.5  \\
J2139+2227    & 06 Jul 2016 & 5 & 4.3  \\
                        & 02 Aug 2016 & 10 & 5.3  \\
                        & 03 Aug 2016 & 10 & 4.5  \\
                        & 04 Aug 2016 & 10 & 7.2  \\
                        & 05 Aug 2016 & 15 & 6.9  \\
                        & 08 Aug 2016 & 5 & 2.8  
\enddata
\end{deluxetable}

We observed each of these targets with the ProEM camera on the McDonald Observatory 2.1-m Otto Struve Telescope.  The ProEM camera is a frame-transfer CCD that obtains time series photometry with effectively zero readout time.  The CCD has 1024$\times$1024 pixels and a field of view of $1.6'\times 1.6'$.  We bin 4$\times$4 for an effective plate scale of 0.36$''$\,pix$^{-1}$. All observations were made through a 3\,{mm} BG40 filter, which blocks light redward of $\approx 6500$\,\AA\ to reduce the sky background.  A complete journal from 31 nights of observing these stars is provided in Table~\ref{tab:obs}.

We obtained at least 31 dark frames of equal exposure time as our science frames, as well as dome flat field frames, at the start of each night. 

\section{Analysis}
\label{sec:anal}

For each run, we measure circular aperture photometry in the dark-subtracted, flatfielded frames for the target and nearby comparison stars with the {\sc iraf} package {\sc ccd\_hsp}, which relies on tasks from {\sc phot} \citep{Kanaan2002}.  We use the {\sc wqed} software \citep{WQED} to divide the target counts by the summed counts from available comparison stars to remove the effect of variable seeing and transparency conditions during each observing run.  {\sc wqed} also applies a barycentric correction to our timestamps to account for the light travel time to our targets changing as the Earth moves in its orbit.  

We search for significant signals of astrophysical variability in the resultant light curves.  We present individual analyses for each target below, sorted into three groups by our ultimate classification of the objects: new pulsating stars, binary systems with photometric variability related to their orbits, and systems for which we can only put limits on a lack of photometric variability.

\subsection{Pulsating Stars}

Most pulsating stars, including most WD pulsators, oscillate at multiple simultaneous frequencies.  We find multiple significant, independent frequencies of photometric variability in three of our targets: SDSS\,J1735+2134, SDSS\,J2139+2227 and SDSS\,J1355+1956.

\subsubsection{SDSS\,J1735+2134}
\label{sec:j1735}

\begin{figure}[b]
  \centering
  \includegraphics[width=1\columnwidth]{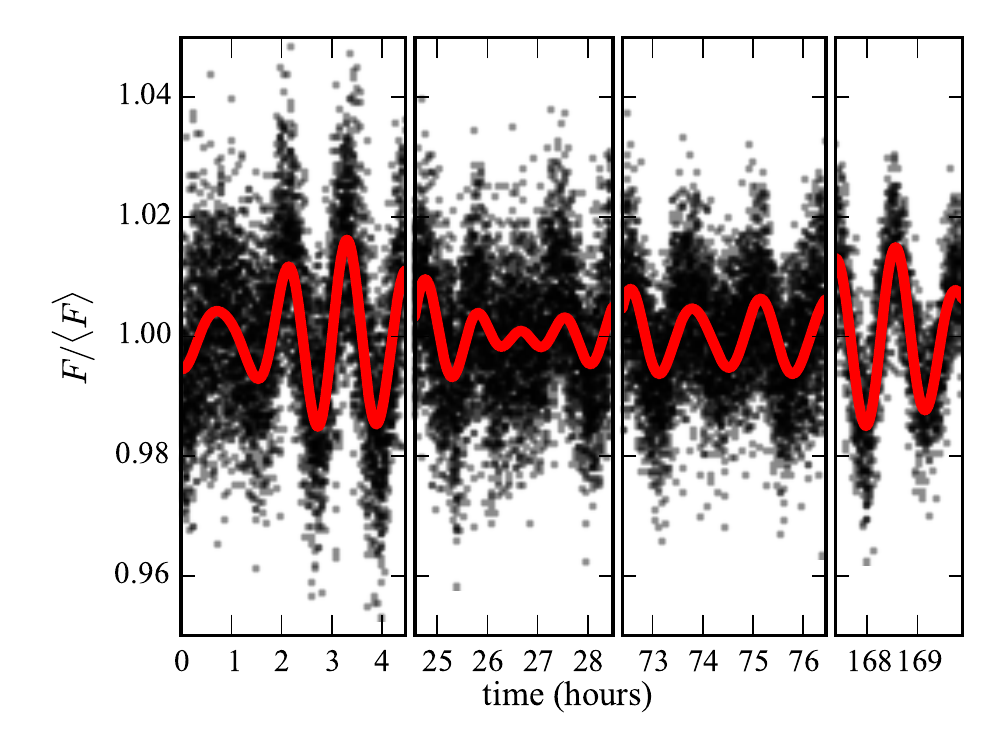}
  \caption{The light curves of SDSS\,J1735+2134 from 4 nights are displayed in black dots.  The x-axis units are hours since the start of the 30 Apr 2016 run. The y-axis gives the ratio of the measured flux relative to the mean flux.  Our 4-period model fit to the data is displayed as a solid line.}
  \label{fig:j1735lc}
\end{figure}

We observed SDSS\,J1735+2134 over 4 nights between 30 Apr and 07 May 2016.  These light curves, displayed in Figure~\ref{fig:j1735lc}, evidence multi-periodic pulsations reaching up to 3\% peak-to-peak amplitude.

We take an iterative approach to determining the pulsation properties of this target.  To detect a new mode, we calculate the Fourier transform (FT) of the combined light curve and assess whether the highest peak exceeds an adopted $4\langle A\rangle$ significance threshold, where $\langle A\rangle$ is the mean amplitude in a local 1000\,$\mu$Hz region of the FT \citep[this corresponds to $\approx$99.9\% confidence;][]{Breger1993,Kushnig1997}. If a significant signal is present, we find the non-linear least-squares fit of a sinusoid to the data, using the peak amplitude and frequency from the FT as initial guesses.  We then ``prewhiten'' the light curve by subtracting out this best fit and compute the FT of the residuals.  If another significant signal is detected above $4\langle A\rangle$ in the FT of the residuals, we redo the non-linear fit with a sum of sinusoids.  We repeat this process until no new significant signals are found.

For SDSS\,J1735+2134, we find 4 significant signals corresponding to 4 eigenfrequencies of this pulsating star.  Their properties are collected in Table~\ref{tab:j1735}, along with analytical uncertainties \citep{Montgomery1999}.  We use millimodulation amplitude (mma) as our unit for pulsation amplitude, where 1 mma = 0.1\% flux variation.

\begin{figure}[t]
  \centering
  \includegraphics[width=1\columnwidth]{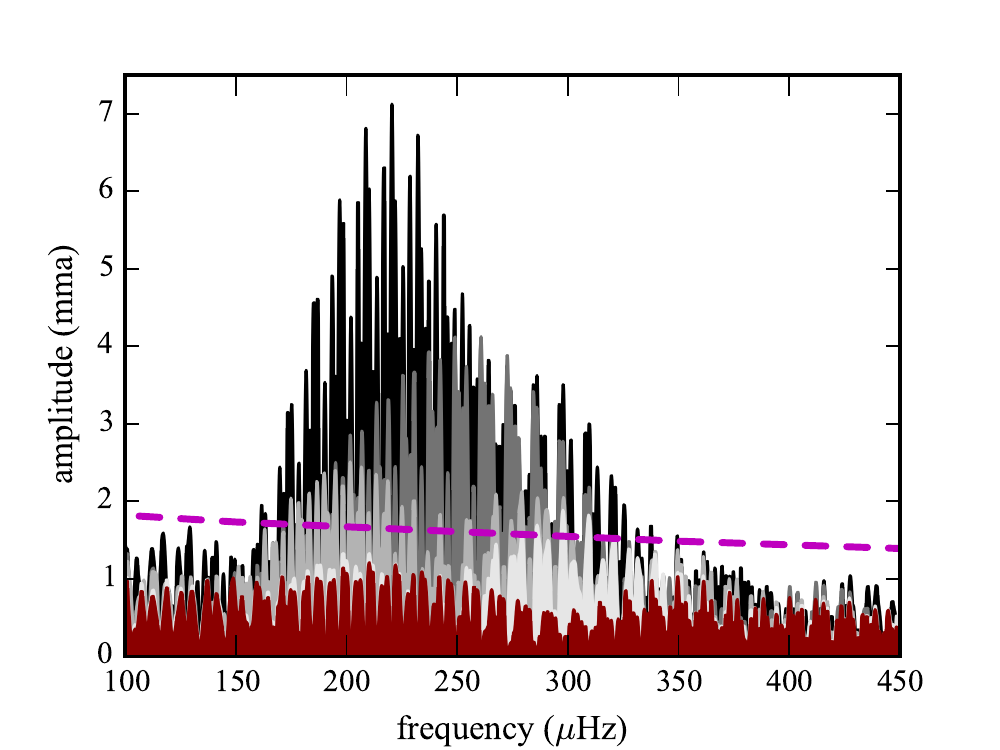}
  \caption{Fourier transforms of the original (black) light curves, increasingly prewhitened data (lighter shades of gray; see text), and final residuals (red) for SDSS\,J1735+2134 in the region of significant pulsational power.  The dashed line is the final $4\langle A\rangle$ significance threshold for the data prewhitened by the four sinusoids characterized in Table~\ref{tab:j1735}.}
  \label{fig:j1735ft}
\end{figure}

The sequence of FTs corresponding to all iterations of our mode detection algorithm is displayed in Figure~\ref{fig:j1735ft}.  The original FT is in black, with increasingly lighter shades of gray representing the FTs of the prewhitened light curves after additional mode detections. The red FT is of the fully prewhitened data and the dashed line is the final $4\langle A\rangle$ significance threshold.

\begin{deluxetable}{c c c c}[!b]
\tablecolumns{4}
\tablecaption{Pulsation Properties of SDSS\,J1735+2134 \label{tab:j1735}}
\tablehead{
\colhead{Mode} & \colhead{Frequency} & \colhead{Period} & \colhead{Amplitude} \\
\colhead{} & \colhead{($\mu$Hz)} & \colhead{(min)}& \colhead{(mma)} }
\startdata
$f_{1}$  & 220.172(0.013) &  75.698(0.004)  &  7.60(0.11)\\
$f_{2}$  & 260.79(0.03) &  63.909(0.007)  &  3.64(0.11)\\
$f_{3}$  & 201.56(0.03) &  82.687(0.012)  &  3.38(0.11)\\
$f_{4}$  & 297.38(0.05) &  56.046(0.009)  &  2.04(0.11)
\enddata
\end{deluxetable}

\subsubsection{SDSS\,J2139+2227}

\begin{figure}[t]
  \centering
  \includegraphics[width=1\columnwidth]{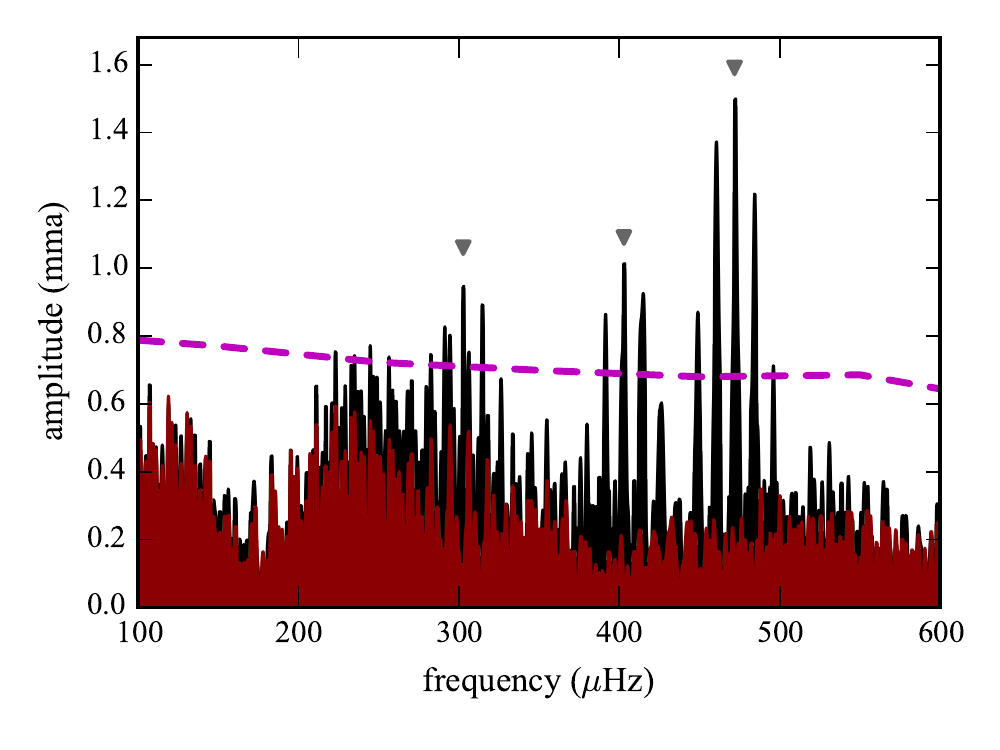}
  \caption{Fourier transform of the original (black) and fully prewhitened (red) light curves of SDSS\,J2139+2227 covering the full region of significant pulsational power.  The dashed line is the final $4\langle A\rangle$ significance threshold for the data prewhitened by the three sinusoids characterized in Table~\ref{tab:j2139} and indicated here with triangles.}
  \label{fig:j2139ft}
\end{figure}

We characterize the pulsations of SDSS\,J2139+2227 from 26 hours of photometry obtained over the span of 7 nights in early Aug 2016\footnote{One nearby comparison star in the field of view, SDSS\,J213905.27+222709.1 ($g=$ 16.77 mag),  was incidentally observed to show deep eclipses while we were monitoring SDSS\,J2139+2227.  The eclipses last $\approx$3 hours and decrease the flux in the $BG40$ filter by $\approx$16\% at mid eclipse.  We observed similar eclipses 1.848 days apart, but the binary period could be an integer fraction of that.}.  The same iterative FT, least-squares fitting, and prewhitening process as used for the previous object reveals the three significant pulsation frequencies that are listed in Table~\ref{tab:j2139}.  The FT before and after prewhitening is displayed in Figure~\ref{fig:j2139ft}.  The pulsation amplitudes are too small relative to the photometric signal-to-noise ratio to be clearly apparent to the eye in the light curve.

\begin{deluxetable}{c c c c}[!b]
\tablecolumns{4}
\tablecaption{Pulsation Properties of SDSS\,J2139+2227 \label{tab:j2139}}
\tablehead{
\colhead{Mode} & \colhead{Frequency} & \colhead{Period} & \colhead{Amplitude} \\
\colhead{} & \colhead{($\mu$Hz)} & \colhead{(min)}& \colhead{(mma)} }
\startdata
$f_{1}$  & 471.82(0.06) &  35.324(0.004)  &  1.52(0.08)\\
$f_{2}$  & 402.85(0.09) &  41.372(0.009)  &  1.02(0.08)\\
$f_{3}$  & 302.73(0.09) &  55.055(0.016)  &  0.99(0.08)
\enddata
\end{deluxetable}

\subsubsection{SDSS\,J1355+1956}

\begin{figure}[t]
  \centering
  \includegraphics[width=0.99\columnwidth]{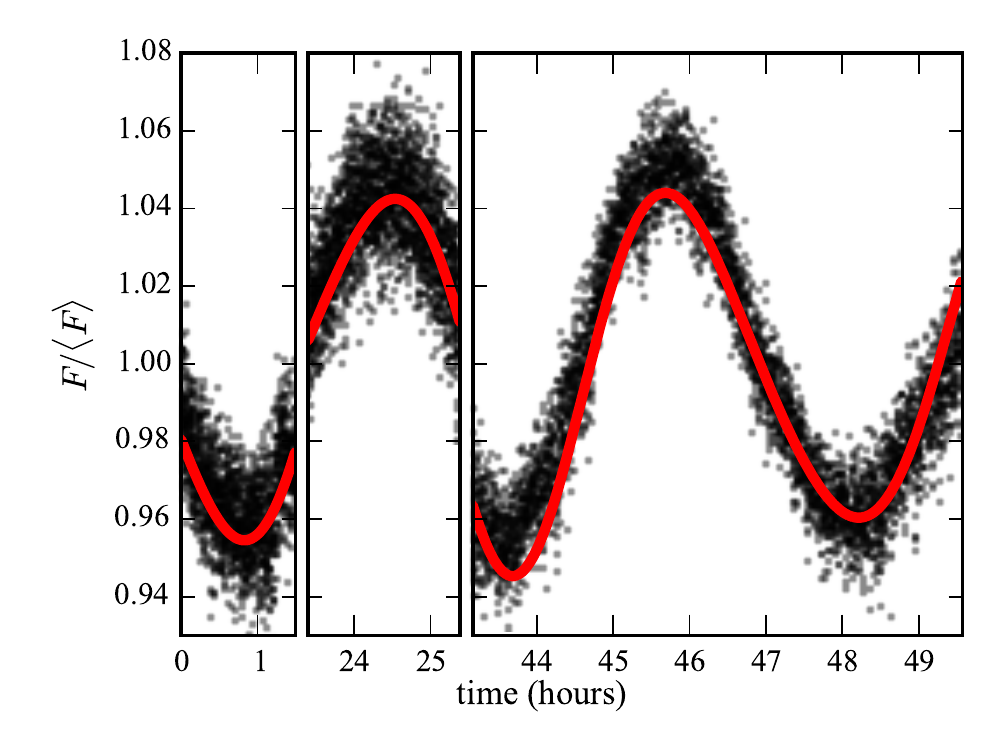}
  \caption{The light curves of SDSS\,J1355+1956 from three consecutive nights in May 2016.  The best-fit two-sinusoid model is plotted over the data.}
  \label{fig:j1355}
\end{figure}

The target SDSS\,J1355+1956 shows a dominant signal with such a long period that only our 6.41\,hr run from 06 May 2016 captured a full cycle.  Figure~\ref{fig:j1355} displays the light curves that we obtained on three consecutive nights, 04--06 May 2016.  Since the durations of the earliest two runs are shorter than the dominant period, they suffer some non-ideal normalization in our standard reduction pipeline.  To account for this, we fit multiplicative scaling factors to the different May 2016 runs simultaneously with the least-squares sinusoid-fitting step of our period search algorithm for renormalization.

\begin{figure}[b]
  \centering
  \includegraphics[width=0.99\columnwidth]{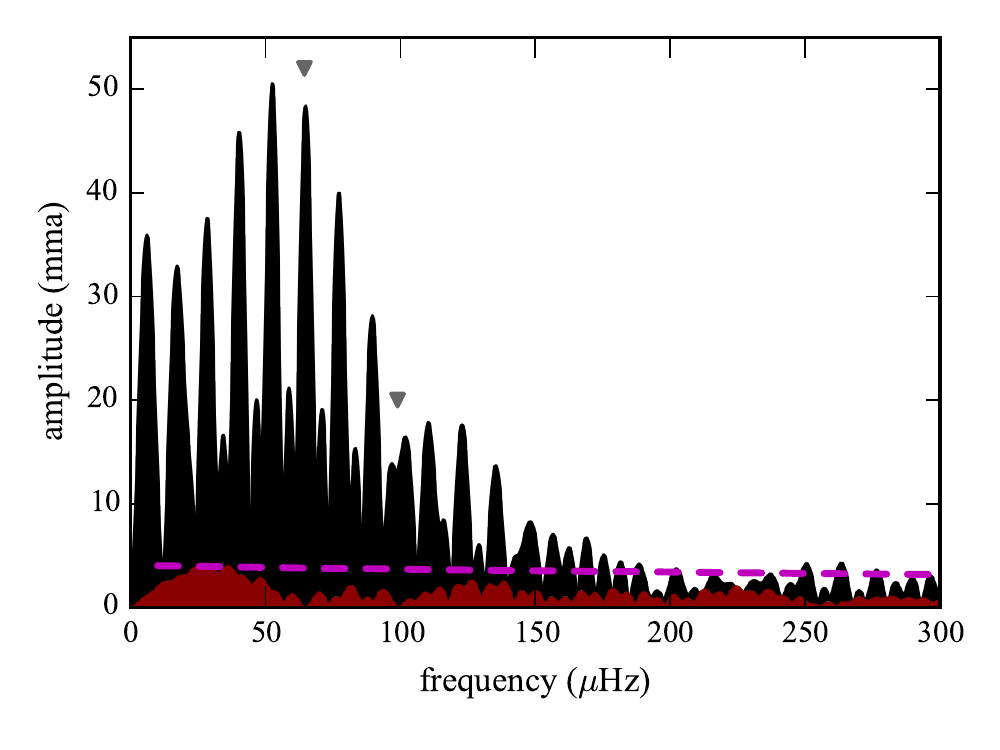}
  \caption{The Fourier transform of the scaled original (black) and fully prewhitened (red) light curves of SDSS\,J1355+1956 from May 2016.  The dashed line shows the final $4\langle A\rangle$ significance threshold for the prewhitened light curve. The frequencies of the two significant modes in Table~\ref{tab:j1355} are marked with triangles.}
  \label{fig:j1355ft}
\end{figure}

The FT of the 06 May 2016 run alone provided an initial guess of 4.74\,hr for the dominant period; however, this value aligns poorly with the data from the two previous nights.  The Catalina Sky Survey \citep[CSS;][]{Drake2009} Data Release 2\footnote{\url{http://nesssi.cacr.caltech.edu/DataRelease/}} provides 321 epochs of well calibrated photometry from eight seasons of observations that we use to guide our mode selection from the complicated alias structure in the FT of our May 2016 data (Figure~\ref{fig:j1355ft}). Rather than the highest peak in the FT of our data (corresponding to $5.2604\pm0.0011$ hours), the CSS data prefer a period near 4.29 hours. We use this as an initial guess in calculating the least-squares single-sinusoid fit to the three-night light curve (with free renormalization parameters).  The FT of the residuals supports the presence of a second significant frequency in this star.  A simultaneous fit of two sinusoids to the data gives our final solution, with parameters listed in Table~\ref{tab:j1355}. This solution is plotted over the observed light curves in Figure~\ref{fig:j1355}.

\begin{deluxetable}{c c c c}[!b]
\tablecolumns{4}
\tablecaption{Pulsation Properties of SDSS\,J1355+1956 \label{tab:j1355}}
\tablehead{
\colhead{Mode} & \colhead{Frequency} & \colhead{Period} & \colhead{Amplitude} \\
\colhead{} & \colhead{($\mu$Hz)} & \colhead{(hr)}& \colhead{(mma)} }
\startdata
$f_{1}$  &  64.430(0.010) &  4.3113(0.0007)  &  46.18(0.16)\\
$f_{2}$  & 98.94(0.05) &  2.8075(0.0015)  & 8.94(0.16)
\enddata
\end{deluxetable}

Figure~\ref{fig:j1355ft} includes the FTs of the rescaled light curve both before and after prewhitening our two-period solution.  The dominant period exceeds the theoretical limit for pulsations in ELM WDs as discussed in Section~\ref{sec:disc}.  The residuals are just barely shy of our adopted significance criterion at a period of 7.295 hours. A pulsation mode of this duration could account for the apparent residual disagreement found in the last panel of  Figure~\ref{fig:j1355}, though this could also be attributed to differential extinction between the target and comparison stars during this long run (see Section~\ref{sec:limits}).

\subsection{Photometric Binaries}

Binary systems can be photometrically variable for many reasons: primary and secondary eclipses, ellipsoidal variations (tidal distortion), reflection, and relativistic Doppler beaming.  We detect photometric variability related to the binary orbital periods determined from RV variations (see Table~\ref{tab:objs}) in two of our targets.

\subsubsection{WD\,J0308+5140}

WD\,J0308+5140 is the only target that we observed that does not fall within the SDSS footprint; it was instead originally identified from a LAMOST \citep[Large Sky Area Multi-Object Spectroscopy Telescope;][]{Wang1996,Cui2012} spectrum.  For convenience, we follow the convention of \citet{ELM6} and include it in tables under the ``SDSS'' column header.   

This target shows the longest-period RV variations in our sample at 19.342$\pm$0.009\,{h}.  Our data reveal dramatic photometric variability related to the orbit in partial coverage of the binary period.

\begin{figure}[b]
  \centering
  \includegraphics[width=0.99\columnwidth]{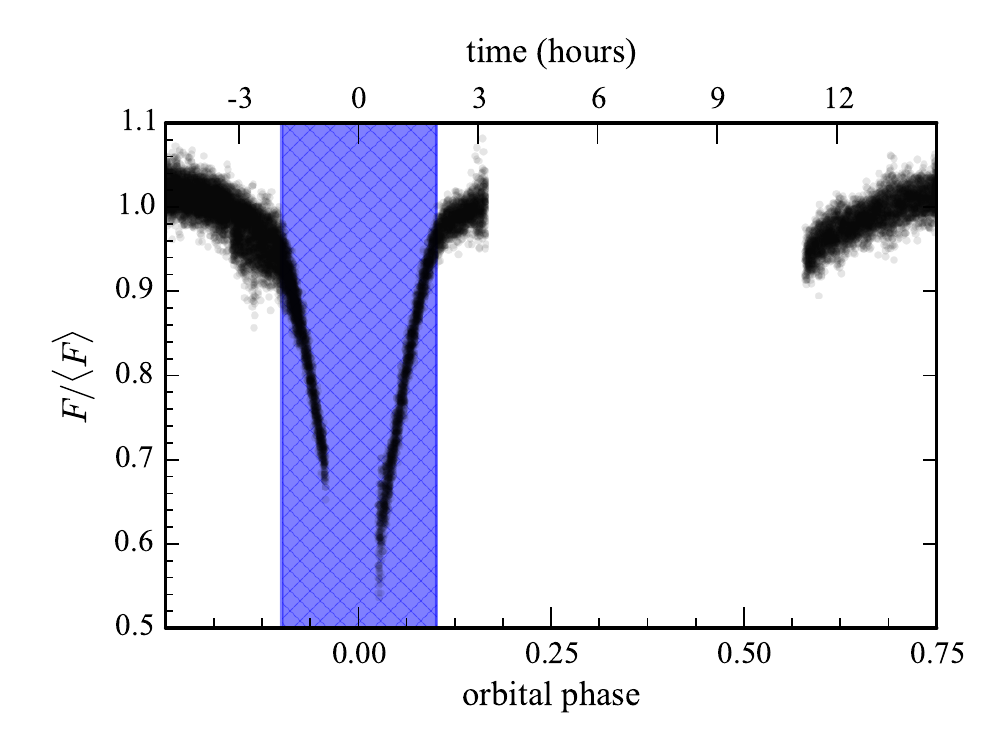}
  \caption{The phase-folded light curve of WD J0308+5140 shows evidence of eclipses that implies a primary star radius $\gtrsim 0.4$ \rsun.  The blue hatched region marks the observed eclipse.}
  \label{fig:0308}
\end{figure}

Figure~\ref{fig:0308} displays the light curve folded on the measured orbital period.  We normalized the target counts summed in the aperture for WD\,J0308+5140 by those of a single, similarly bright ($B$ = 16.5 mag) field comparison star---entry 1350-03091578 from USNO-A2.0 located at $\rm{RA}(2000)=03^{\rm{h}} 08^{\rm{m}} 19\fs87$, $\rm{Dec.}(2000)=51\degr 40\arcmin 34\farcs 15$ \citep{Monet1998}---so that the individual runs align smoothly.

While our phase coverage is not complete enough to precisely determine system parameters, we identify by eye the apparent start and end times of a deep primary eclipse. This range, centered on phase 0, is highlighted in Figure~\ref{fig:0308}.

We rely on five simplifying assumptions to calculate a lower limit on the radius of the primary star: (1) the stars are in circular orbits; (2) the relative velocity between the two stars equals the measured RV semi-amplitude of $K_1 = 78.9\pm 2.7$ km s$^{-1}$; (3) the catalogued $K_1$ value represents only the speed of the primary star; (4) the system inclination is $90\degr$; and (5) the two binary components have equal radii. Under this oversimplified model, the radius of the primary star is related to the measured eclipse duration, $\Delta t$, by the expression $R_1 = K_1\Delta t / 4$.  With an eclipse duration of $\approx$4 hours, we have $R_1 \gtrsim 0.4$ \rsun . The first assumption is supported by the sinusoidal fit to the RV measurements in ELM6. If any of the latter four assumptions are false, we would find a larger radius for the primary star, so our result is a conservative lower limit.

\begin{figure}[b]
  \centering
  \includegraphics[width=0.99\columnwidth]{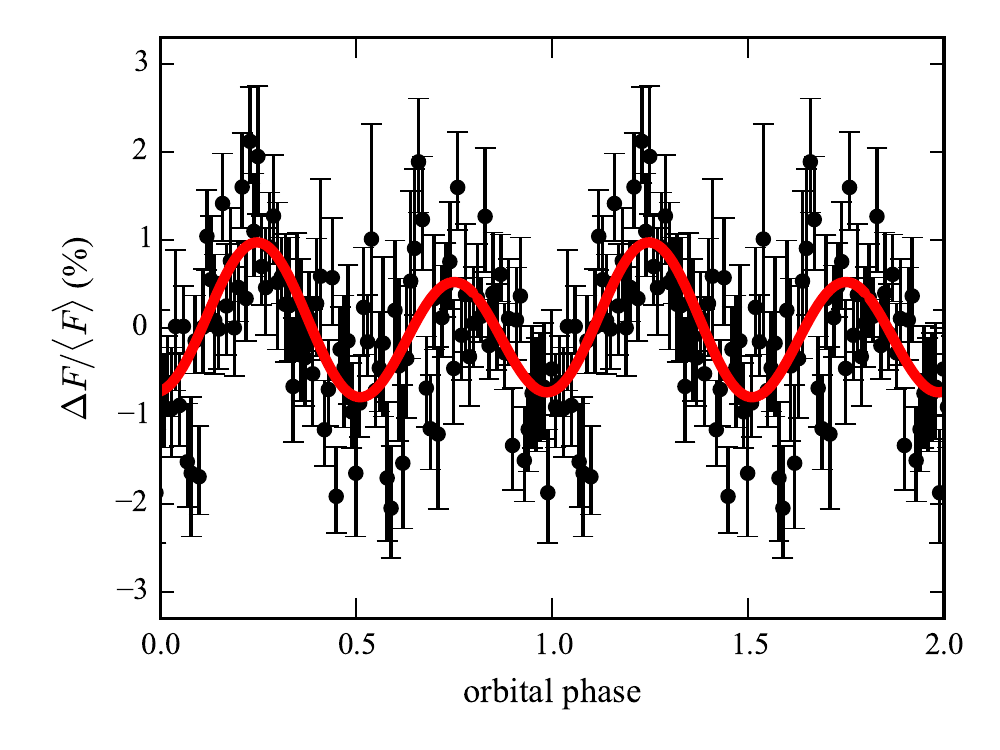}
  \caption{The phase-folded, binned light curve of SDSS J1054$-$2121 shows evidence of ellipsoidal variations and relativistic Doppler beaming. Our best fit model is plotted over the data in red.}
  \label{fig:1054}
\end{figure}

\subsubsection{SDSS\,J1054$-$2121}

While we see no evidence of pulsational variability in the light curve of SDSS J1054$-$2121, it does show photometric variability related to the binary orbital period of $2.51 \pm 0.16$\,{h} determined from RV measurements.  

Because of the long gap between our short March 2015 run and our 7.28\,{h} of data that April, we use only the April data in this analysis.  Since both April runs exceed one full orbital cycle, we divide a straight line fit from each light curve to correct for differential extinction effects without concern for missing longer-timescale variations. 

The FT of these data reveals a dominant signal at $1.251\pm 0.004$\,{h} (with additional extrinsic uncertainty of $\pm 0.07$\,h from the aliasing structure of the spectral window) consistent with half the orbital period. We interpret this as the signature of tidally induced ellipsoidal variations of the star.

\begin{figure*}[t]
\centering
\includegraphics[width=1.99\columnwidth]{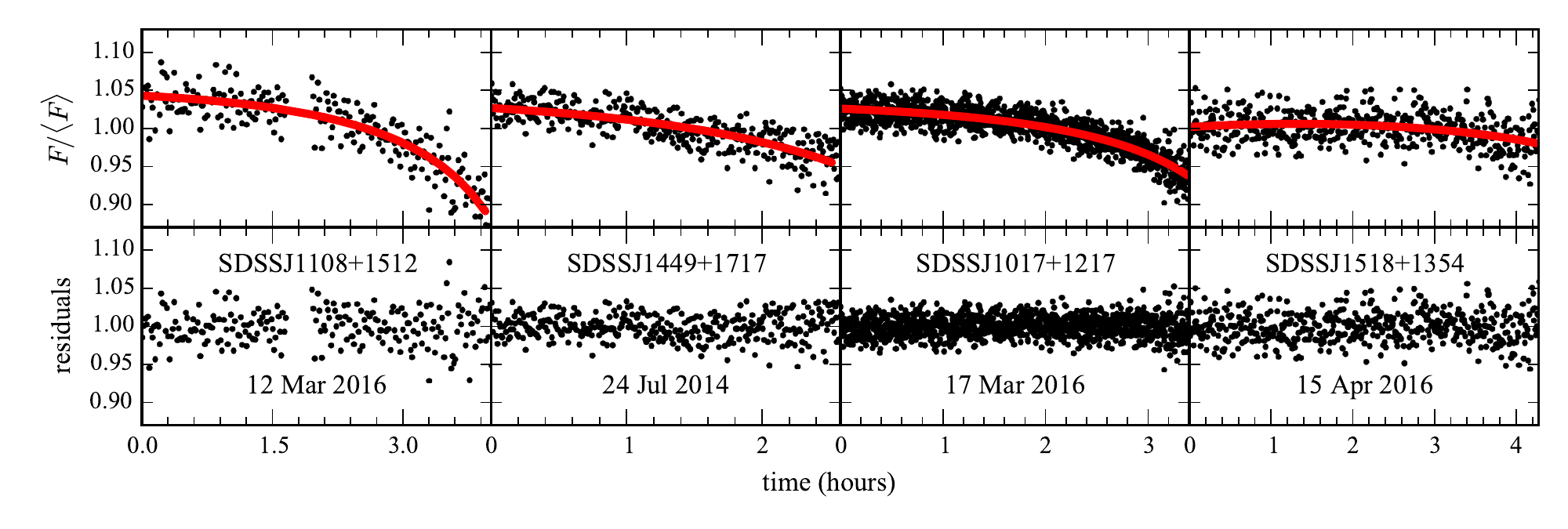}
  \caption{Each plot in the top panel displays the divided light curve with the largest overall trend for each of the four targets that do not show clear signs of intrinsic stellar variability. The solid red lines show the best-fit differential extinction model for each.  Our analysis of the residuals displayed in the bottom panel reveals no significant astrophysical signal.}
\label{fig:airmass}
\end{figure*}

We phase-fold the April data on the refined binary period and then average the photometry within 100 phase bins, each having width 1.5\,{min} and containing 7--16 measurements.  We calculate the standard deviation of points within each phase bin and divide that by the square root of the number of points to get error bars for the binned, phase-folded light curve.  This light curve is repeated through two full orbital cycles in Figure~\ref{fig:1054}.

The dominant sinusoidal signal is from ellipsoidal variations, which has peaks twice per orbit when the elongated side of the tidally distorted ELM is presented to our line of sight.  We refer to this as the $\cos{2\phi}$  term with angular frequency 2 cycles orbit$^{-1}$---where phase zero ($\phi = 0$) is defined as when the ELM WD is furthest from us.

Doppler beaming is the modulation of the measured flux with the radial velocity of the target, caused by both the Dopper shift of flux in/out the observational bandpass and the relativistic beaming of light in the direction of motion \citep[e.g.,][]{RL,vanKerkwijk2010}. With our phase convention, this has a $\sin{\phi}$ behavior with frequency 1 cycle orbit$^{-1}$.  The amplitude of this effect is directly related to the RV semi-amplitude of the target through $A_{\rm DB} = -(3-\alpha)K_1/c$, where $\alpha = d\log{F_\nu}/d\log{\nu}$ is the spectral index, which accounts for the Dopper shift of flux into the observational bandpass.  We estimate the spectral index of SDSS J1054$-$2121 to be $\alpha = 0.956$ by averaging the mean $\alpha$ for our best-fit model spectrum in each of the two wavelength ranges 3200--3600\,\AA\ and 5500--6500\,\AA , which correspond approximately to the blue and red edges of the BG40 bandpass.  With an RV semi-amplitude of  $261.1\pm 7.1$\,km\,s$^{-1}$, we expect to measure a Doppler beaming signal of $\approx 0.18$\% in this system.

\begin{deluxetable}{l c c c}[t]
\tablecolumns{9}
\tablecaption{Least-squares Amplitudes \label{tab:sinamps}}
\tablehead{
\colhead{SDSS} & \colhead{$\cos{2\phi}$ (\%)} &  \colhead{$\sin{\phi}$ (\%)}  & \colhead{$\cos{\phi}$ (\%)} } 
\startdata
J1054$-$2121    & 0.75(0.08) & 0.23(0.08) & 0.03(0.08)
\enddata
\end{deluxetable}

A $\cos{\phi}$ component of the light curve could be present from reflection if the ELM WD's companion is sufficiently hot, but \citet{Hermes2014} did not find this effect to a significant level in 20 double-degenerate binaries with low-mass primary stars.

We compute a least-squares fit for the $\cos{2\phi}$, $\sin{\phi}$ and $\cos{\phi}$ amplitudes, along with the phase and an overall vertical offset, to the folded light curve.  Our best-fit model is overplotted in red in Figure~\ref{fig:1054}.  The reduced $\chi^2$ of this five-parameter fit is 0.85.  The amplitudes of the three sinusoidal components are given in Table~\ref{tab:sinamps}.  We calculate the uncertainties from the diagonal elements of the covariance matrix after scaling the photometric uncertainties to give $\chi^2_{\rm red} = 1$.

The $\sin{\phi}$ amplitude is within $1\sigma$ of the expected 0.18\% and the $\cos{\phi}$ term is consistent with zero. The $\cos{2\phi}$ term is entirely consistent with ellipsoidal variations in an ELM WD.  A more thorough analysis of this target, including a refinement of system parameters from the photometric data, will be presented in follow-up work.

\subsection{Null Results}
\label{sec:limits}

For the remaining four targets of the present survey, we do not detect significant astrophysical signals in our data.  However, the extent of our observational coverage is not sufficient to completely rule out photometric variability in these stars.  Since stellar pulsations and orbital timescales can be on the order of hours for ELM WDs, we are careful not to classify a star as a nonvariable without multiple individual runs of at least this duration.  We are cautious because multiple sources of variability (e.g., two pulsation modes) can happen to destructively combine during an individual night's observations, masking the signal. Sky and transparency conditions also commonly vary on timescales of hours and can leave signatures in the data.

For some observing runs on our remaining targets, we do see overall long-timescale trends throughout the divided light curves.  This is likely due to differential extinction with changing airmass during a night's observations.  Since the spectral energy of our targets is generally distributed differently (usually more toward shorter wavelengths) across the observational bandpass than nearby comparison stars, light from the target will experience a different amount (usually more) of atmospheric scattering on the way to our detector.

For normal-mass WDs, where pulsation periods of $\sim $10 minutes are usually much shorter than the duration of observations, we typically mitigate this effect by fitting and dividing out a low-order polynomial \citep[e.g.,][]{Nather1990}.  However, when searching for signals with timescales on the order of the run duration, this approach is inappropriate as it may mistakenly remove astrophysical signals of interest.

\begin{deluxetable}{c c c c}[t]
\tablecolumns{4}
\tablecaption{Limits on Pulsations in NOVs \label{tab:limits}}
\tablehead{
\colhead{SDSS} & \colhead{Period} &or& \colhead{Amplitude} \\
\colhead{} & \colhead{(h)}&&  \colhead{(mma)} }
\startdata
J1108+1512  & $> 4.0$ & &  $< 13.4$ \\
J1449+1717  & $> 2.6$ & & $< 9.8$ \\
J1017+1217  & $> 3.9$ & & $< 5.3$ \\
J1518+1354  & \ldots & & \ldots 
\enddata
\end{deluxetable}

Instead, we divide from each light curve the least-squares fit of $a\exp{bX}$, where $X$ is the airmass at each frame and $a$ and $b$ are free coefficients. This approach will not represent differential extinction well if there are major changes in atmospheric conditions during observations or if extinction has an azimuthal dependence at the observing site, but this first-order approach appears to fully explain the dominant trends found in the light curves of our remaining targets.

The top panels of Figure~\ref{fig:airmass} display the light curves with the most pronounced airmass trend for each remaining target (from left to right): SDSS\,J1108+1512, SDSS J1449+1717, SDSS\,J1017+1217 and SDSS\,J1518+1354.  The solid lines are the least-squares fits of the differential extinction models, and the bottom panels show the final reduced light curves after dividing out these systematics.  FTs of these fully reduced light curves, and those from all other runs on these targets, do not reveal significant signals to our $4\langle A\rangle$ significance threshold (see Section~\ref{sec:j1735}).

We place conservative limits on possible pulsation amplitudes and periods that may be present in these objects in Table~\ref{tab:limits}.  Since we impose a careful requirement of considering at least two multi-hour light curves before designating a star as not observed to vary (NOV), we do not provide limits for SDSS\,J1518+1354, the only target in our sample that was observed on only one night.  For the others, we base our quoted limits on the two longest light curves for each object alone, claiming that no pulsations are present with periods shorter than the second-longest observing run and amplitudes greater than the largest $4\langle A\rangle$ threshold value in the FT of either run.

It is worth noting that a peak in the FT of the combined runs on SDSS\,J1017+1217 from 30 Apr and 03 May 2016 exceeds a lower $3\langle A\rangle$ level, and we consider this feature with period $48.569\pm 0.006$\,min and amplitude $2.7\pm 0.5$\,mma suggestive.

Additional observations of any of these targets could reveal lower amplitude or longer timescale variations.

\begin{figure*}[t]
\centering
\includegraphics[width=1.9\columnwidth]{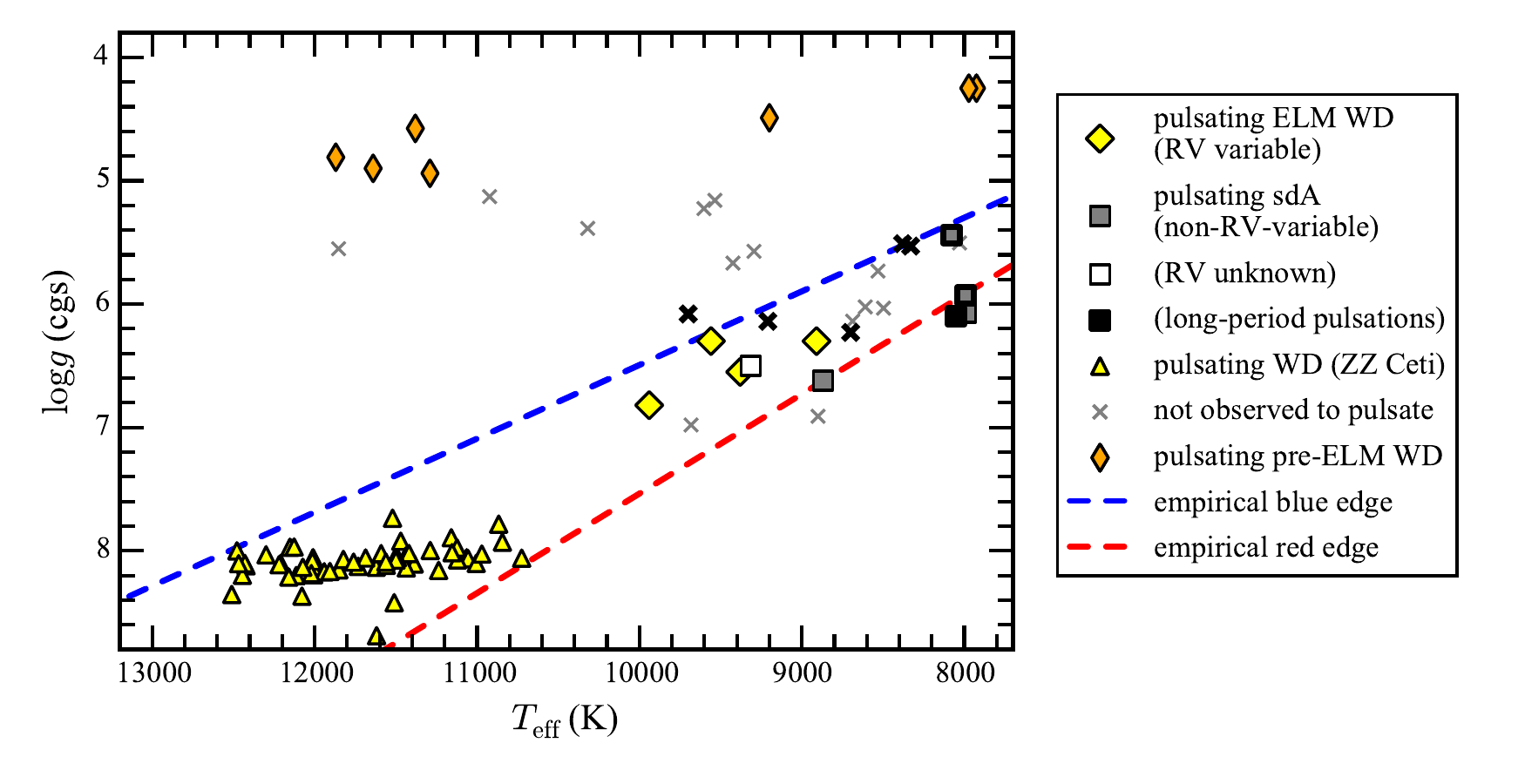}
  \caption{The locations of known pulsating stars with ELM-like spectra in \logg --\teff\ space.  Pulsating ELMVs confirmed with RV variations \citep{ELMV1,ELMV23,ELMV7} are indicated with yellow diamond markers.  Pulsating ELM WD candidates without measured RV variations are marked with squares, including three objects from this work and the targets published in \citet{ELMV45} and \citet{ELMV6}.  The filled black square represents SDSS\,J1355+1956, which cannot be a WD, and the white square is SDSS J1618+3854 \citep{ELMV6} that has not yet been observed with time series spectroscopy. Objects with constraints on a lack of pulsations from time series photometry are marked with $\times$ symbols \citep[this work;][]{ELMV1,ELMV23,ELMV45}.  Pulsating pre-ELM WDs from \citet{Maxted2013,Maxted2014,Corti2016} and \citet{Gianninas2016} are marked with orange narrow diamonds.  Typical \logg\ $\sim 8$ ZZ Ceti pulsators from \citet{Gianninas2011}, corrected for 3D convection effects \citep{Tremblay2013}, are marked with yellow triangles. The empirical DA instability strip published in \citet{ELM6} is marked with dashed lines.  The objects presented in this work are outlined with thicker black borders.}
\label{fig:loggteff}
\end{figure*}

\section{Discussion}
\label{sec:disc}

In our search for photometric variability from nine candidate pulsating ELM WDs, we identified significant signals in five targets.  However, the observed properties of some of these targets are not in agreement with the ELM WD classification.

Since ELM WDs can only form through mass transfer in close binary systems, we expect to be able to measure orbital RV variations for these stars, except in very few nearly face-on ($i \lesssim 20\degr$) cases. \citet[][{Section 3.4}]{ELM7} determine that the total 11 non-RV-variable objects (8 with \teff\ $< 9{,}000$ K) out of 78 targets with $\log{g}< 7.15$ catalogued in the ELM Survey likely represents an overabundance to a 2.5$\sigma$ significance compared with expectations from a random distribution of orbital orientations.  This suggests that some of these non-RV-variable objects may not be bona fide ELM WDs.

For one of the non-RV-variable ELM WD candidates, SDSS\,J1355+1956, we measure an exceptionally long dominant pulsation period of $4.3113\pm0.0007$ hours. Following \citet{Hansen1985}, we calculate the approximate theoretical maximum allowed nonradial gravity mode pulsation period of $P_{\rm max}\approx 45$\,min for a WD with the published spectroscopic parameters of this target, assuming an Eddington gray atmosphere.  The observed pulsations greatly exceed this theoretical limit for surface reflection in a WD, providing additional evidence that this star is individually a false positive in the ELM Survey. This strongly supports that SDSS\,J1355+1956 is not a WD, and its actual surface gravity is likely less than the spectroscopically determined value \logg\ $= 6.10\pm0.06$. With the dominant mode amplitude reaching 41.51 mma, we are likely observing pressure-mode pulsations in a high-amplitude $\delta$ Scuti---a class of pulsating star typically found in the range $6000 \lesssim {T}_{\mathrm{eff}} \lesssim 9000$ K \citep[e.g.,][]{Uytterhoeven2011}. However, recent analysis of the hot, lead-rich subdwarf UVO 0825+15 by \citet{Jeffrey2016} provides compelling evidence for pulsation periods that exceed the \citet{Hansen1985} limit, casting some doubt on the robustness of this theoretical result. Given the large amplitude and the upper limit on RV semi-amplitude from ELM7 of $K_1 < 40.9$\,km s$^{-1}$, the observed variability cannot be attributed to ellipsoidal variations of an ELM WD \citep{Morris1993}.

Of the remaining pulsating candidate ELM WD variables, only four out of eight show RV variations in time series spectroscopy: SDSS J184037.78+642312.3 \citep{ELMV1}, SDSS J111215.82+111745.0, SDSS J151826.68+065813.2 \citep{ELMV23}, and PSR J1738+0333 \citep{ELMV7}. The two other new pulsating stars described in this work are not RV variables, as is the case for the previously published pulsators SDSS J161431.28+191219.4 and SDSS J222859.93+362359.6 \citep{ELMV45}.  It is unknown whether another claimed ELMV---SDSS J161831.69+385415.15 \citep{ELMV6} that was identified as an ELM candidate from SDSS spectroscopy \citep{Kepler2015}---is RV variable.  We submit that none of these non-RV-variable pulsating stars have been conclusively shown to be ELMVs.  Some could be in nearly face-on binary systems, but when we simulate random binary orientations, we find the probability of four out of eight systems with $i < 20\degr$ to be $< 0.0008$.

\citet{Kepler2016} found thousands of objects with SDSS DR12 spectra that are consistent with ELM WDs that they call ``sdAs,'' with the ELM classification requiring confirmation of RV variations. The sdAs are strongly concentrated around \teff\ $\approx 8000$ K, which is where the DA WD instability strip extends through the ELM regime.  There is no evolutionary scenario that predicts such an abundance of ELM WDs at this temperature, which may highlight an inaccuracy in current spectroscopic models or their application.  We suspect that SDSS\,J1355+1956 and some of our other non-RV-variable pulsating stars are actually members of this sdA class.  This does imply that the sdAs also pulsate in or near the same region of spectroscopic parameter space, revealing the potential for distinguishing between sdAs and ELM WDs asteroseismically.

We depict the present landscape of WD pulsations in \logg --\teff\ space in Figure~\ref{fig:loggteff}.  We distinguish confirmed ELMVs (yellow diamonds) from pulsating candidate ELM WDs without measured RV variations (squares). The black square corresponds to SDSS\,J1355+1956, with a much longer pulsation period than expected from an ELM WD. The white square is SDSS J161831.69+385415.15 \citep{ELMV6}, which does not have available time series spectroscopy. The symbols representing objects analyzed in this work are outlined with bold black borders.   We include NOVs with limits on pulsational variability ($\times$), more massive ZZ Ceti variables (triangles), and pulsating pre-ELMs (orange narrow diamonds) for context.  The empirical bounds of the DA instability strip from \citet{ELM6} are marked with dashed lines.  If we redefine these boundaries based on only the confirmed ELMVs, we find a more narrow extension of the strip to low \logg .

We also observe variability from photometric binaries in our sample.  Partial coverage of the 19.342-hour binary period of WD J0308+5140 reveals evidence of eclipses.  The lower limit on the primary star radius of $R_1 \gtrsim 0.4$ \rsun\ is inconsistently large compared with the maximum expected radius for a cooling ELM WD. The evolutionary models of \citet{Althaus2013} give a maximum cooling-track radius of  $\approx$0.13 \rsun , while the \citet{Istrate2016} models find a maximum of $\approx$0.17 \rsun .  However, the models with element diffusion enabled show that some ELM WDs can temporarily become much larger during CNO flashes as they settle onto their final cooling tracks \citep[and previous works referenced therein]{Althaus2013,Istrate2016}.  \emph{Kepler} photometry of the eclipsing system KIC 10657664 has demonstrated empirically that ELM WDs can be at least as large as $0.15\pm0.01$ \rsun\ \citep{Carter2011}. Additional photometry of WD J0308+5140 would provide some of the first precise constraints on the physical properties of sdA stars. 

The presence of this false positive in the ELM Survey cautions that binary confirmation alone is not sufficient to positively identify an ELM WD.  The properties of WD\,J0308+5140 are similar to another eclipsing system, SDSS J160036.83+272117.8, which was not included in the ELM6 sample due to eclipse durations that were inconsistent with the ELM WD classification \citep[][2017 in prep.]{Wilson2015}.  Only binary RV periods short enough to preclude non-degenerate stellar components ($P_{\rm orb} \lesssim 6$\,hr), or those with supporting data as photometric binaries, should be interpreted as ELM WDs with confidence.

The other binary that we observe photometric variations of, SDSS\,J1054$-$2121, is just such a case.  The ellipsoidal variation signature of $0.75\pm0.08$\% amplitude is entirely consistent with that expected for a double-degenerate binary with an ELM WD primary.  In future work, we will use the measured ellipsoidal variability amplitude to significantly improve our physical constraints on this system.

\section{Summary and Conclusions}
\label{sec:conc}

We identified nine candidate pulsating ELM WDs from the ELM Survey papers VI and VII.  Each of these targets has spectroscopically determined \teff\ and \logg\ values that place them within 500 K of the empirical low-mass extension of the DA WD instability strip, which overlaps the population of sdA stars with $\langle {T}_{\rm eff}\rangle \approx 8000$ K.  We obtained time series photometry of these systems from McDonald Observatory, most over many nights.

The following are our main results:
\begin{itemize}[noitemsep,leftmargin=*,topsep=0pt]
\item[--] Fourier analysis reveals that three targets---SDSS\,J1355+1956, SDSS\,J1735+2134, and SDSS J2139+2227---show significant pulsational variability. However, since these targets are among the few for which time series spectroscopy from ELM7 did not show the RV variations that are expected from an ELM WD, we do not consider them confirmed ELMVs.
\item[--] In particular, SDSS\,J1355+1956 pulsates with a dominant period of $4.3113\pm0.0007$ hours, far exceeding the theoretical limit for pulsations in a WD.  This is likely a $\delta$ Scuti variable with an overestimated \logg\ from spectroscopic model fits.
\item[--] A total of 4 out of 8 other pulsating variable stars in the parameter space of ELM WDs do not show significant RV variations in time series spectroscopy.  There is less than a 0.0008 probability that these are all nearly face-on ($i < 20\degr$) binaries.  Some of these targets are likely sdA stars---a stellar population revealed in recent SDSS data releases \citep{Kepler2016} of unclear nature.
\item[--] Our data on WD\,J0308+5140 reveal evidence for a deep $\approx$4\,hr eclipse, implying that the primary star has radius $\gtrsim 0.4$ \rsun . This is not consistent with an ELM WD and demonstrates that a mere detection of RV variations is not sufficient to make this classification, though very short period binaries may exclude other classes.
\item[--] Ellipsoidal variation and Doppler beaming amplitudes measured in SDSS J1054$-$2121 are consistent with the ELM WD classification for this object.\\
\end{itemize}

We note that the remaining ambiguity of the nature of the non-RV-variable objects with ELM-like spectral lines will be largely resolved by Gaia astrometric solutions, including for all ELM Survey objects, within the next few years. This will allow us to determine not only the stellar types of individual objects, but also the relative sizes and spatial distributions of the different stellar populations that occupy this region of spectroscopic parameter space.

\acknowledgements We thank the anonymous referee for feedback that improved this manuscript.  K.J.B., D.E.W., M.H.M., B.G.C., K.I.W. and Z.V. acknowledge support from NSF grant AST-1312983.  Support for this work was provided by NASA through Hubble Fellowship grant \#HST-HF2-51357.001-A, awarded by the Space Telescope Science Institute, which is operated by the Association of Universities for Research in Astronomy, Incorporated, under NASA contract NAS5-26555. A.G. and M.K. gratefully acknowledge the support of the NSF under grant AST-1312678. K.J.B. thanks S.~O.~Kepler and Thomas Kupfer for helpful discussions, and John Kuehne for improvements to the 2.1-m Otto Struve Telescope that made observations of this quality possible. This paper includes data taken at The McDonald Observatory of The University of Texas at Austin.  The authors acknowledge the Texas Advanced Computing Center (TACC) at The University of Texas at Austin for providing data archiving resources that have contributed to the research results reported within this paper. The CSS survey is funded by the National Aeronautics and Space
Administration under Grant No. NNG05GF22G issued through the Science
Mission Directorate Near-Earth Objects Observations Program.  The CRTS
survey is supported by the U.S.~National Science Foundation under
grants AST-0909182 and AST-1313422. This research has made use of the VizieR catalogue access tool, CDS, Strasbourg, France.

\end{document}